\documentclass[prb,twocolumn,showpacs,floatfix,superscriptaddress,epsf,psfig]{revtex4}
\usepackage{epsfig}
\usepackage{graphicx}
\usepackage{epstopdf}
\begin{document}

\title{\textit{Ab initio} Investigation on Hybrid Graphite-like Structure Made up of Silicene and Boron Nitride}

\author{C. Kamal}
\affiliation{Indus Synchrotrons Utilization Division, Raja Ramanna Centre for Advanced Technology, Indore 452013, India }
\author{Aparna Chakrabarti}
\affiliation{Indus Synchrotrons Utilization Division, Raja Ramanna Centre for Advanced Technology, Indore 452013, India }
\affiliation{Homi Bhabha National Institute, Raja Ramanna Centre for Advanced Technology, Indore 452013, India}
\author{Arup Banerjee}
\affiliation{Homi Bhabha National Institute, Raja Ramanna Centre for Advanced Technology, Indore 452013, India}
\affiliation{ BARC Training School, Raja Ramanna Centre for Advanced Technology, Indore 452013, India }

\begin{abstract}
In this work, we report our results on the geometric and electronic properties of hybrid graphite-like structure made up of silicene and boron nitride (BN) layers.  We predict from our calculations that this hybrid bulk system, with alternate layers of honeycomb silicene and BN, possesses physical properties similar to those of bulk graphite. We observe that there exists a weak van der Waals interaction between the layers of this hybrid system in contrast to the strong inter-layer covalent bonds present in multi-layers of silicene. Furthermore, our results for the electronic band structure and the density of states show that it is a semi-metal and the dispersion around the Fermi level (E$_F$) is parabolic in nature and thus the charge carriers in this system behave as \textit{Nearly-Free Particle-Like}. These results  indicate that the electronic properties of the hybrid bulk system resemble closely those of bulk graphite. Around E$_F$ the electronic band structures have contributions only from silicene layers and the BN layer act only as a buffer layer in this hybrid system since it does not contribute to the electronic properties near E$_F$.
In case of bi-layers of silicene with a single BN layer kept in between, we observe a linear dispersion around E$_F$ similar to that of graphene. However, the characteristic linear dispersion become parabola-like when the system is subjected to a compression along the transverse direction. Our present calculations show that the hybrid system based on silicon and BN can be a possible candidate for two dimensional layered system akin to graphite and multi-layers of graphene.
\end{abstract}
\maketitle

\section{Introduction}

There has been a lot of interest in layered systems,
such as  two-dimensional (2D) graphene-like honeycomb structure made up of materials other than carbon, because of the novel properties associated with these 2D systems and also due to their potential applications in nanotechnology. The important property associated with the layered systems is that they have a strong in-plane bonding and a weak van der Waals (vdW) bonding in a direction perpendicular to the plane which leads to an anisotropic bonding arrangement in the systems\cite{nicolosi-sci2013}. 
Among these graphene-like systems, \textit{silicene}, the graphene analog of silicon,  have been extensively studied by  both theoreticians and experimentalists in recent past\cite{sili-ciraci1,sili-ciraci2,ck-jpcm2013,ck-crc2013,ni,drummond,eza1,ck-arxive,ck-arxive-optical,eza2,eza3,sili-soc,sili2,sili3,refpap1,refpap5,sili-grown, sili-expt,lay,sili-expt4,sili-expt5,sili-expt6,sili-multi,sinr1,sinr2,gurel,avilal,sili-flake}. Silicene possesses many physical properties which are similar to those of graphene. For example, silicene is a semi-metal and the charge carriers in this 2D material behave like massless Dirac-Fermions due to the presence of linear dispersion around Fermi level (E$_F$) at a symmetry point $K$ in the reciprocal lattice\cite{sili-ciraci1,sili-ciraci2}.  Presence of linear dispersion in silicene has been recently confirmed  by ARPES measurement\cite{lay}. Moreover, it has been observed that a band gap can be opened up and tuned in a monolayer of silicene by applying an external transverse electric field\cite{ck-arxive,ni,drummond,eza1} which is, however, not possible in monolayer of graphene.

Though the properties of monolayer of silicene resemble those of monolayer of graphene, the properties of multi-layers of silicene are drastically different from those of multi-layers of graphene. It has been observed from our recent studies\cite{ck-jpcm2013,ck-crc2013} that the multi-layers of silicene possess strong inter-layer covalent bonds in contrast to the weak van der Waals interaction between the layers of graphene multi-layers and bulk graphite. Presence of inter-layer strong covalent bond influences many properties of multi-layers of silicene. Due to this strong inter-layer covalent bonding, the multi-layers of silicene can no longer behave like a layered system.  
The reason for differences in properties between carbon and silicon based systems is due to the different energetically favourable hybridizations present in these two systems in spite of the fact that these two atoms contain same number of electrons in the valence shell. Most favourable hybridization in silicon system is sp$^3$ whereas carbon system can exist in sp, sp$^2$, as well as sp$^3$ hybridizations. It is important to note that the difference in the values of energy levels between the two valence sub-shells namely 3s and 3p in Si atom is smaller than the corresponding value between 2s and 2p sub-shells in C atom. This leads to the preference of sp$^3$ hybridization in Si because 3s  can easily mix with all the sub-shells 3p$_x$, 3p$_y$ and 3p$_z$.

Due to the above mentioned reason, it is not possible to have bi- and multi-layers of silicene analogous to the bi- and multi-layers of graphene. Furthermore, the graphite-like layered structure of silicon cannot be constructed by directly stacking one silicene layer over another since they readily form inter-layer covalent bonds\cite{ck-jpcm2013,ck-crc2013}. However, it is desirable to obtain graphene-like silicon based layered systems possessing similar exciting and novel properties of multi-layer of graphene since the former has an important advantage over carbon based systems because of their compatibility with the existing semiconductor industry. Keeping these points in mind, in the present work, we propose that there is   a possibility of creating a graphite-like layered structure of silicon by inserting a buffer layer in between the multi-layers of silicene. The buffer layer prevents the strong inter-layer covalent bonds between the layers of silicene.  Moreover, the buffer layer is expected not to alter the electronic properties of the multi-layers of silicene around E$_F$.  For this purpose, we consider a hybrid graphite-like layered system made up of alternate layers of honeycomb silicene and honeycomb boron nitride. It has been shown in existing theoretical studies that interaction between monolayer of silicene and boron nitrate substrate is due to weak van der Waals\cite{si-bn1, si-bn2,si-bn3,si-bn4}.  We study the geometric and electronic structures of the hybrid graphite-like layered system  by employing \textit{ab initio} density functional theory (DFT)\cite{dft} based calculations. In this case, BN layer acts as a buffer layer. We want to probe whether it would be possible to have an energetically stable hybrid system with physical properties similar to those of bi-layers of graphene as well as bulk graphite.

The present paper has been arranged in the following manner. We give the details of the computational methods employed in our calculations in the next section. The results of the properties of hybrid graphite-like structure have been discussed in section III, and then followed by the conclusion in section IV.  Effect of vdW interaction on the geometric and electronic properties of the hybrid system is studied in the appendix.

\section{Computational details}

Density functional theory (DFT)\cite{dft} based calculations have been performed using Vienna ab-initio simulation package (VASP)\cite{vasp} within the framework of the projector augmented wave (PAW) method. We employ generalized gradient approximation (GGA) given by Perdew-Burke-Ernzerhof (PBE)\cite{pbe} for exchange-correlation functional. 
The plane waves are expanded with energy cut of 400 eV. We use Monkhorst-Pack scheme for k-point sampling of  Brillouin zone integrations with 10$\times$10$\times$5  and 11$\times$11$\times$1 for bulk and bi-layer systems respectively.   The convergence criteria for energy in SCF cycles is chosen to be 10$^{-6}$ eV. The geometric structures are optimized by minimizing the forces on individual atoms with the criterion that the total force on each atom is below 10$^{-2}$ eV/ $\AA$.  For bi-layer, we use a super cell geometry with a vacuum of about 15  $\AA$  in the z-direction (direction perpendicular to the plane of silicene/BN) so that the interaction between two adjacent unit cells in the periodic arrangement is negligible. The geometric structures and charge density distributions are plotted using XCrySDen software\cite{xcrysden}.

\section{Results and Discussions}
\subsection{Geometric structures and charge density distributions}

\begin{figure}
 \begin{center}
 \includegraphics[width=8cm,bb=0 170 600 750]{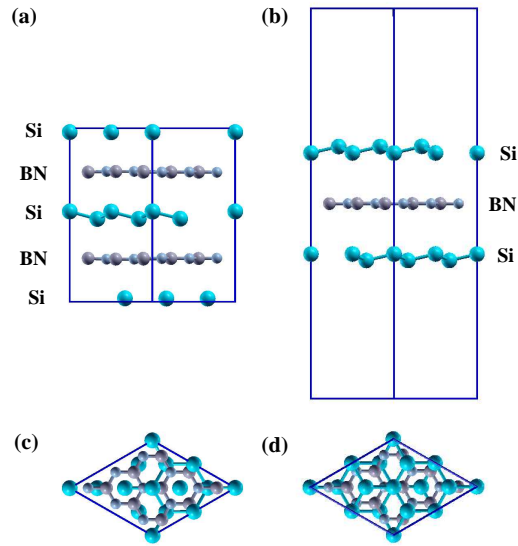}
  \caption{ (color on line) The optimized geometric structures of hybrid graphite-like structures made up of silicene and boron nitride layers. Bulk system (Si$_{16}$B$_{18}$N$_{18}$) : (a) side and (c) top views. Bilayer of silicene with a single boron nitride layer (Si$_{16}$B$_{9}$N$_{9}$) : (b) side and (d) top views.}
 \end{center}
  \end{figure}

 \begin{figure}
 \begin{center}
 \includegraphics[width=8cm,bb=0 170 900 800]{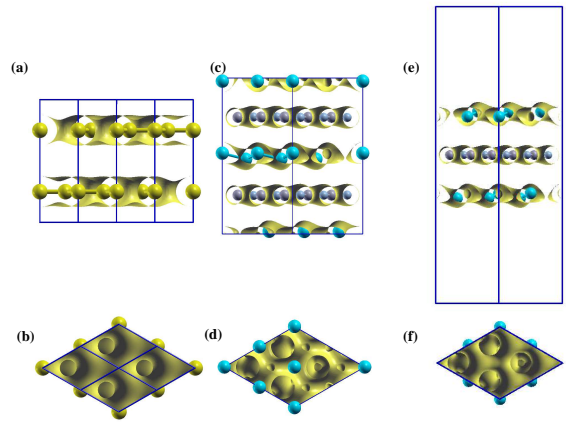}
  \caption{ (color on line) Spatial charge density distributions of layered systems. Bulk graphite with 2$\times$2$\times$1 super cell ( (a) side and (b) top views) and hybrid graphite-like structures made up of silicene and boron nitride layers: bulk system (Si$_{16}$B$_{18}$N$_{18}$) ((c) side and (d) top views) and bi-layer (Si$_{16}$B$_{9}$N$_{9}$)  ((e) side and (f) top views) }
 \end{center}
  \end{figure}

In this section, we start our discussion on the geometric properties of the hybrid graphite-like bulk system (Si$_{16}$B$_{18}$N$_{18}$) with alternate layers of honeycomb silicene and BN. The space group of  the hybrid bulk system is  P3m1. The unit cell contains two silicene and two BN (buffer) layers. The super cell of (2$\times$2) of honeycomb silicene is lattice matched with 3$\times$3 of honeycomb BN with the deviation of  only about 1.7 $\%$. Each layer of silicene (BN) contains 8 silicon (9 boron and 9 nitrogen) atoms. The calculated optimized geometric structures of the hybrid graphite-like structure are shown in Fig. 1 (a) and (c). Our results on geometric structure obtained by DFT based calculation with PBE XC functional show that the hybrid system has a value of lattice constant about 7.554 $\AA$ along $a$ axis (which is denoted by $A$). The calculated values of Si-Si and B-N bond lengths in the basal plane are 2.246 and 1.454 $\AA$ respectively. We observe that the amount of buckling present in silicene layer in the hybrid system is slightly increased to 0.543 $\AA$ from its free standing value\cite{ck-jpcm2013} of 0.457 $\AA$. The reason for increase in the buckling length is due to the interaction of silicene layer with the other layers present in the bulk system. Moreover, the increase in the value of buckling leads to a higher contribution of sp$^3$-like hybridization in Si atoms in silicene of the hybrid system as compared to that of Si atoms in free standing silicene. Our calculations with PBE XC functional give 14.695 $\AA$ for the value of lattice constant along $c$ axis (which is denoted by $C$). Thus, the inter-layer distance, denoted by $D$, between the silicene and BN layers becomes 3.674 $\AA$ (in this case, $D= C/4$,  See Fig. 1(a) ). We also perform similar calculations for bi-layer of silicene with a single BN layer kept in between (Si$_{16}$B$_{9}$N$_{9}$).  The optimized geometry of bi-layer is given in Fig. 1 (b) and (d). In this case, the lattice constant '$A$' and inter-layer distance of the hybrid bi-layer are estimated to be 7.607 and 4.011 $\AA$ respectively. We observe from these results that for hybrid bi-layer both the values of lattice constant and inter-layer distance slightly increase from the corresponding values in hybrid bulk system. Furthermore, our calculations of cohesive energy of the hybrid systems show that they are energetically stable. We find that the cohesive energy per atom for the hybrid bulk and bi-layer systems are 6.07 and 5.57 eV/atom respectively. These results indicate that the interaction between the layers in bilayer is slightly weaker as compared to that in the hybrid bulk.

In order to get deeper insight into the interaction between the layers of the hybrid systems, we plot the spatial distribution of charge densities for the hybrid graphite-like bulk system (Si$_{16}$B$_{18}$N$_{18}$) (see (c) and (d)) and bi-layer of silicene (Si$_{16}$B$_{9}$N$_{9}$) (see (e) and (f)) in Fig. 2. These figures clearly show the charge distribution of hybrid systems exhibit characteristic features of layered systems such as (i) presence of more charge densities in basal planes containing covalent bonds between atoms in each layer and (ii) negligible amount of charge densities present between the adjacent layers. We also compare these results with that of bulk graphite (see (a) and (b)) and find that the charge density distributions of  the hybrid systems are very similar to that of bulk graphite. Thus, our results clearly indicate that the hybrid systems are similar to those of graphite and multi-layers of graphene. 

We also determine the equation of state (EOS)) of the hybrid bulk system (Si$_{16}$B$_{18}$N$_{18}$). In Fig. 3, we provide the results for variation in total energy of system with the volume of unit cell. We evaluate the bulk modulus of the hybrid system by fitting our results given in Fig. 3 with Birch-Murnaghan formula for EOS. From the fitting,  the values of bulk modulus and its derivative with respect to pressure for this system are estimated to be 48.5 GPa and 1.76 respectively.  This value of bulk modulus of hybrid bulk system is slightly higher than the corresponding values of layered structures such as bulk graphite (33.8 GPa)\cite{bulk-graphite} and bulk hexagonal BN (25.6 GPa)\cite{bulk-bn}. However, they are much lower than that of bulk silicon (98 GPa)\cite{bulk-si}. This result again corroborates with our earlier results which indicates the layered nature of the bulk hybrid system made up of silicene and BN.  Thus, the hybrid system may also be used as a soft material similar to graphite and bulk BN.

\begin{figure}
 \begin{center}
 \includegraphics[width=7.5cm]{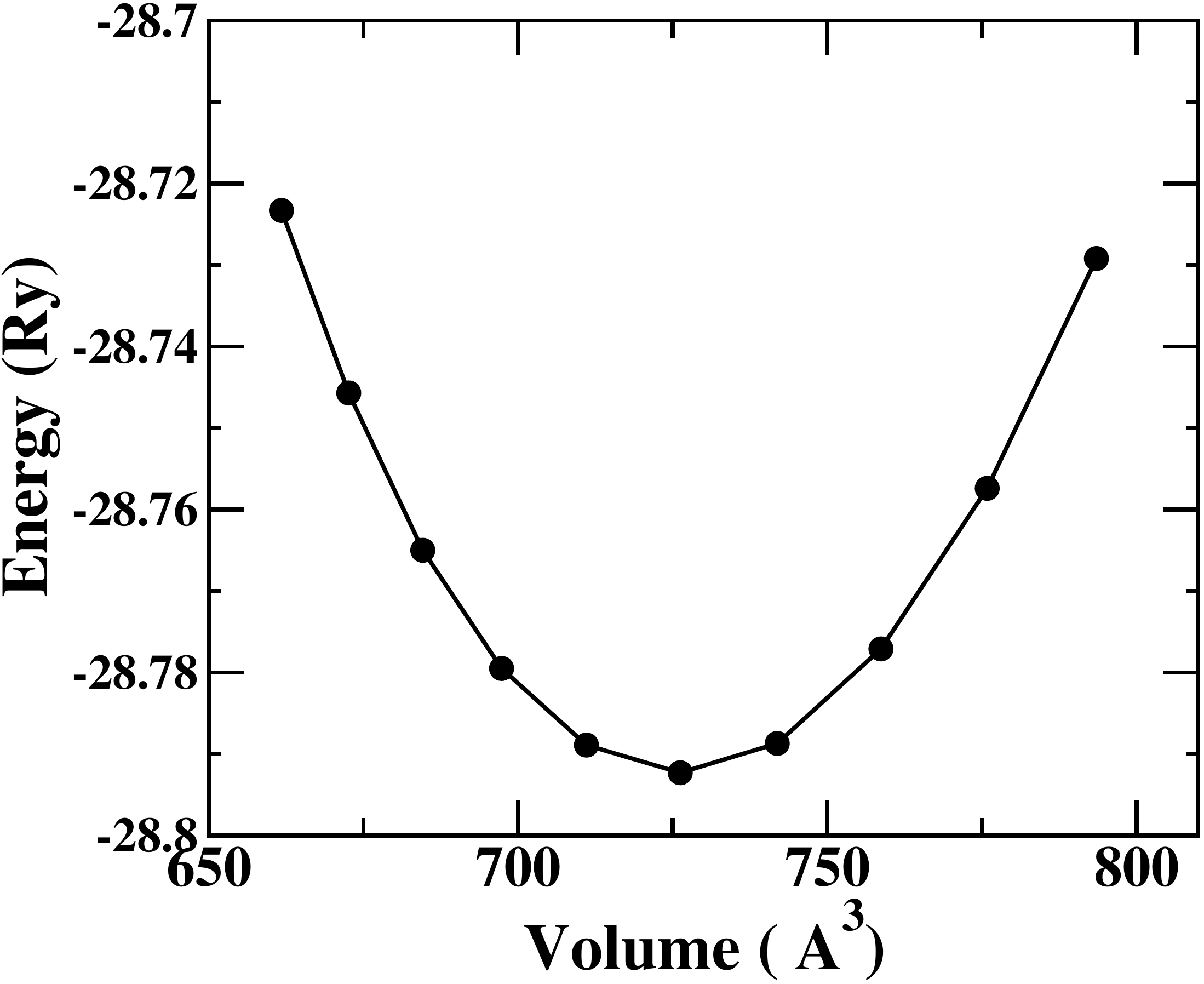}
  \caption{ Variation in total energy of hybrid graphite-like bulk system made up of silicene and BN layers  (Si$_{16}$B$_{18}$N$_{18}$) with its volume.}
 \end{center}
  \end{figure}
\subsection{Electronic Properties}

 \begin{figure}
 \begin{center}
 \includegraphics[width=7.5cm]{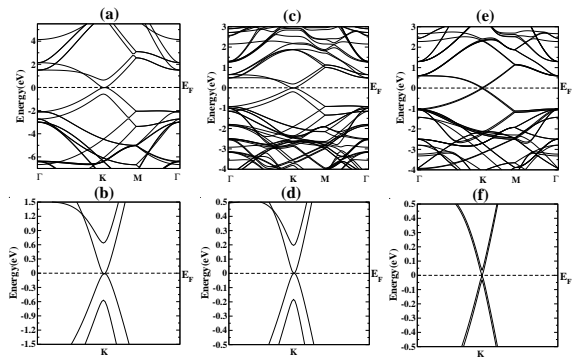}
  \caption{ Band structures of layered systems in two energy ranges. Bulk graphite with 2x2x1 super cell ( (a) and (b)) and hybrid graphite-like structures made up of silicene and boron nitride layers: bulk system (Si$_{16}$B$_{18}$N$_{18}$) ((c) and (d) ) and bi-layer (Si$_{16}$B$_{9}$N$_{9}$)  ((e) and (f) ). The values of energy of bands are with respect to the respective Fermi level.  }
 \end{center}
  \end{figure}

 \begin{figure}
 \begin{center}
 \includegraphics[width=7.5cm]{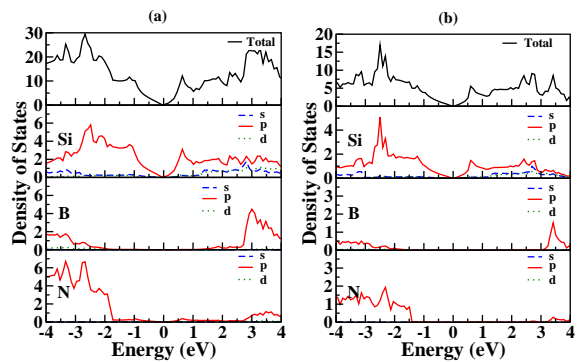}
  \caption{ Total and partial density of states for hybrid graphite-like structure made up of silicene and boron nitride layer: (a) bulk system (Si$_{16}$B$_{18}$N$_{18}$)  and (b) bi-layer (Si$_{16}$B$_{9}$N$_{9}$). The values of energy are with respect to the Fermi level.  }
 \end{center}
  \end{figure}

Having analyzed the geometrical properties of hybrid graphite-like structures in the previous section, we now discuss our results on the electronic properties of these hybrid systems. In Fig. 4, we plot the electronic band structures, in two different energy ranges, for both the hybrid bulk and bi-layered systems along the high symmetry points in reciprocal lattice which give dispersions corresponding to the motion of the charge particles along the planar directions. For the sake of comparison, we also include the electronic band structure of bulk graphite in Fig. 4 (a) and (b). We calculate the band structure for graphite super cell 2$\times$2$\times$1 (containing 16 carbon atoms) so that we can directly compare these results with the hybrid system. Our results on band structure for the hybrid bulk system show that the conduction and valence bands touch each other only at the highly symmetric point $K$ in Brillouin zone. On comparison of  the results for the hybrid bulk system, Si$_{16}$B$_{18}$N$_{18}$, (Fig. 4(c) and (d)) with those of bulk graphite (Fig. 4(a) and (b)), we find that the electronic band structures of these two systems are similar. However, we observe that the hybrid system contains  additional number of bands which lie above $\approx$2.0 eV and below $\approx$-2.0 eV.  The contributions for these bands are mainly due to the BN layers. Furthermore, around the Fermi level, we observe a parabola-like dispersion for the hybrid bulk system (see Fig. 4 (d)) as in graphite. Thus, the charge carriers in this system behave \textit{nearly-free-particle-like}. It is clearly seen from Fig. 4 (b) and (d) that the band structure of the hybrid bulk system and bulk graphite are rather similar close to the Fermi level. This leads to an important conclusion that the hybrid bulk system made up of alternate silicene and BN layers can be a possible material for silicon based layered structure similar to that of carbon based bulk graphite.

Interestingly, in case of the hybrid bi-layered system, we observe a linear dispersion around the Fermi level in contrast to the parabola-like dispersion present in bi-layer of graphene. Furthermore, this dispersion is also distinctly different from the corresponding dispersion for the pure bi-layer (without BN layer) of silicene\cite{ck-jpcm2013} where the parabolic dispersions are shifted in both E and k direction in the band structure due to strong inter-layer covalent bonding. A closer look at the bands reveals that two linear dispersions are present in the band structure and they correspond to the two silicene layers of hybrid bi-layer. This suggests that there exists a much weaker interaction between the layers in bi-layer as compared to that between the layers in bulk system. This result again corroborates with our data on the geometric properties where it is observed that the inter-layer distance in bi-layer is about 9 $\%$ larger as compared to that in the bulk.

In Fig. 5, we present the results of  calculated total density of states (DOS) and partial DOS for these two hybrid systems, namely bulk system (Si$_{16}$B$_{18}$N$_{18}$) and bi-layer (Si$_{16}$B$_{9}$N$_{9}$).  First important observation is that the hybrid systems, both bulk as well as bi-layer, are semi-metallic since the values of DOS at E$_F$ are zero. Investigation from the atom projected partial DOS clearly indicates that the contributions of DOS just below and above Fermi levels are mainly due to $\pi$ and $\pi^*$ orbitals of silicene layers. Moreover, we observe that there is no contribution from either boron or nitrogen (and thus BN layer) around the Fermi level. These atoms contribute to DOS at about -1.8 eV ( and below) and 2.5 eV (and above) compared to the Fermi level respectively for valence  and conduction bands. 

 \begin{figure}
 \begin{center}
 \includegraphics[width=7.5cm]{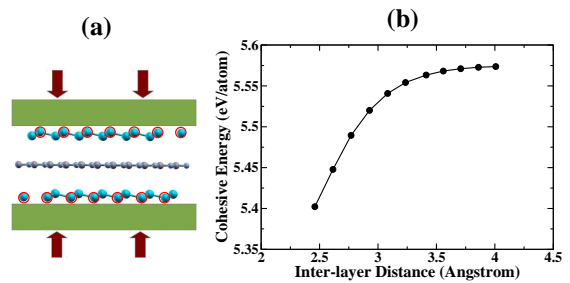}
  \caption{ (color on line)  (a) Hybrid bi-layer of silicene with a single BN layer (Si$_{16}$B$_{9}$N$_{9}$) under  compression along the transverse direction. The z-components of positions of atoms with circle (Red color) are frozen during the geometric optimization. (b) Variation of cohesive energy per atom with inter-layer separation. }
 \end{center}
  \end{figure}

\subsubsection{Variation of Properties with Inter-layer Distance} 
\begin{figure*}
 \begin{center}
 \includegraphics[width=16.cm]{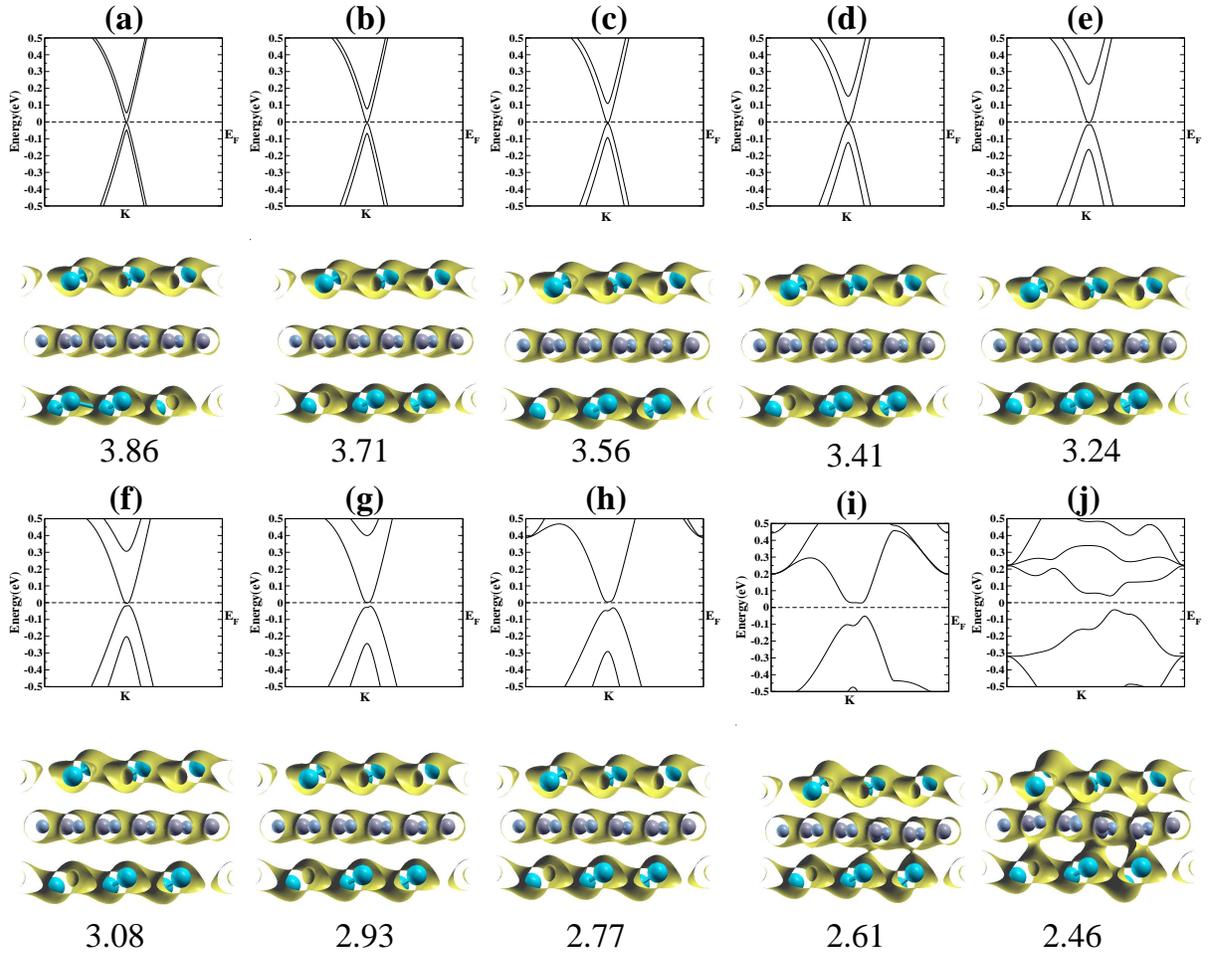}
  \caption{ (color on line) Variation in the electronic band structure and the spatial charge density distribution with different inter-layer separation. The energy of bands are with respect to Fermi level. The value of inter-layer distance (in $\AA$) is given below each figure.}
 \end{center}
  \end{figure*}

In order to understand the effect of inter-layer distance on the properties of the hybrid system, we perform the calculations of electronic band structure of the hybrid bi-layer system under various transverse stress. The compression along the direction perpendicular to the plane of silicene sheets ($c$ axis) is simulated by freezing z-components of the positions of the surface atoms during the geometric structure optimization. These surface atoms are denoted by circle with red color in Fig. 6 (a). The strength of the interaction between the layers is varied by changing the inter-layer distance. The stability of this system under the influence of transverse compression is analyzed by the cohesive energy calculation. Our calculations show that the hybrid bi-layered system under the transverse compression is energetically stable. From Fig. 6(b), we observe that the  value of cohesive energy per atom decreases monotonically with decrease in the value of inter-layer distance.  Investigation on geometric structure indicates that the variation in the value of lattice constant ($A$) due to the compression is very small and maximum variation is less than 1.5 $\%$. However, the buckling length in silicene layer is largely increased, up to 0.884 $\AA$ (for the highest compression)  from 0.543 $\AA$, during the compression. This is due to the increased interaction between the silicene and BN layers.

Now, we discuss the results of band structure and charge density distribution for different inter-layer distances which are given in Fig. 7. 
As the inter-layer distance is decreased the degeneracy of the bands corresponding to the two silicene layer is lifted due to an increased interaction between the layers. It is clearly seen from Fig. 7 that the four bands ( two valence and two conduction bands) of silicene layers become non-degenerate. The characteristics of these four bands decide the electronic properties of bilayer.  Fig. 7 (d)-(f) shows that when the value of  inter-layer distance is little more than 3 $\AA$  the characteristic linear dispersion present in the band structure of this hybrid bi-layer is changed to parabola-like dispersion which are very similar to those of bi-layer of graphene\cite{grap-bi-layers1,grap-bi-layers2,grap-bi-layers3,grap-bi-layers4,grap-bi-layers6,grap-bi-layers7, grap-bi-layers-disp}. Thus, we infer from these results that the weak interaction between the layers changes the character of charge carriers in this 2D system from a Dirac-Fermion to a \textit{Nearly-Free-Particle-Like}.  Furthermore, we also observe that the lower and upper most bands start to move away from the Fermi level as the inter-layer separation is reduced. The corresponding charge density plots for the system with the inter-layer distance (Fig. 7 (d)-(f)) show that the interaction between the layers is still a weak vdW since there is no significant overlap of charges from the atoms present in the adjacent layers. However, when the value of inter-layer distance is about 3 $\AA$ and below, we observe opening of a band gap in the band structure of this hybrid system. Thus, the results of our calculation show that there is a semi-metal to semi-conductor transition due to compression along the direction perpendicular to the plane of silicene sheets ( $c$ axis). Up to this point, we observe no deformation in the geometric structure of the bi-layered system. When the inter-layer distance (Fig. 7(g)-(i)) is further reduced, the interaction between the layer increases which in turn increases the value of induced band gap of the system. We also observe that a significant amount of overlap in the charge density between the layers which indicates the onset of covalent-like bond formation between the atoms in the adjacent layers (See Fig. 7(i)). 

On further reduction in the inter-layer distance to about 2.46 $\AA$, we see from Fig. 7(j) that there is a noticeable deformation in the geometric structure of the hybrid bi-layer. There is an undulation in the geometry of BN layer. One part of the BN layer is moved upward and another goes downward such that B and N atoms present in these two parts make a covalent-like bond with Si atoms of the adjacent silicene layer. The presence of strong covalent-like bonds is confirmed by the charge density analysis (Fig. 7(j)). Again due to these strong bonds, the electronic behavior of the system becomes completely different from that given in Fig. 7(a)-(f)). In this case, we also observe that the system has a direct band gap of about 824 meV.   Furthermore, there is no signature of any parabola-like dispersion around the Fermi level. This clearly indicates that the system no longer behaves like a layered graphite-like material.

\section{Conclusion}
We have carried out studies on the geometric and electronic properties of hybrid graphite-like structures made up of silicene and boron nitride (BN) layers by employing \textit{ab initio} DFT based calculations. We observe from the cohesive energy calculations that the hybrid system is energetically stable. Our calculations predict that the hybrid bulk system possesses physical properties similar to those of bulk graphite. The coupling between the layers of silicene and BN of this hybrid system is due to weak van der Waals interaction which is same as that in graphite and multi-layers of graphene.  We observe from the results on the electronic band structure and the density of states that the hybrid bulk system is a semi-metal and it possesses the dispersion curve, around E$_F$, very similar to that of bulk graphite.  Main contributions to the electronic band structure around E$_F$ arise only due to silicene layers.  Our calculations on bi-layer of silicene with a BN layer show that it possesses the characteristic linear dispersion around E$_F$ due to much weaker interaction between the layers.  However, the nature of dispersion curve becomes parabola-like when the system is compressed along the direction perpendicular to the plane of silicene sheet. Finally, we also observe an opening up of band gap near the Fermi level when the inter-layer distance is about 3 $\AA$ and below.  These calculations show that the hybrid system based on silicon and BN can be a possible candidate for two dimensional layered soft material akin to graphite and multi-layers of graphene. These results would be important from application point of view since the silicon based systems are more compatible with the existing semiconductor industry as compared to the carbon based systems.

\section{Appendix}


\begin{table}
 \caption{The results of optimized geometries of hybrid graphite-like structures made up of silicene and boron nitride obtained by DFT with PBE and vdW-DF exchange-correlation functionals.}
\begin{scriptsize}

\begin{tabular}{|c|c|c|c|c|c|c|}
 \hline
\hline
	&	\multicolumn{4}{|c|}{Lattice Constants  ($\AA$)} & \multicolumn{2}{|c|}{Inter-layer Distance  ($\AA$)}\\  \cline{2-5}
System	      &	\multicolumn{2}{|c|}{$A$}	&	\multicolumn{2}{|c|}{$C$} & \multicolumn{2}{|c|}{$D$}\\ \cline{2-7}
                &   PBE     & vdW-DF  &   PBE     & vdW-DF   &   PBE     & vdW-DF \\
\hline

Bulk 	&	7.554	&	7.539		&	14.695	&	13.704 & 3.674 &3.426\\
(Si$_{16}$B$_{18}$N$_{18}$) 	&		&			&		&	& &\\
\hline
Bi-layer  &	7.607	&	7.646		&	8.021		&	7.903	&4.011 & 3.951\\
(Si$_{16}$B$_{9}$N$_{9}$) 	&		&			&		&	& &\\
\hline 
\hline
\end{tabular}
      \end{scriptsize}
\end{table}

\textbf{Influence of weak van der Waals Interaction}: We would like to mention here that the standard exchange-correlation (XC) functionals such as local density approximation (LDA) and generalized gradient approximation (GGA) do not take into account weak attractive potential arising due to vdW interaction\cite{vdw-df,vdw-df2,vdw-df3,vdw-df4,gga-failure1,gga-failure2,gga-failure3,gga-failure4,gga-failure5,gga-failure6,gga-failure7}. In order to incorporate the weak vdW interaction between the layers of the hybrid system and study its effect on the geometric and electronic properties of the hybrid system, we perform the electronic structure calculations based on DFT with van der Waals density functional (vdW-DF)\cite{vdw-df,vdw-df2,vdw-df3} as implemented in the Quantum-Espresso package\cite{qe}. 
The values of lattice constants obtained by vdW-DF are summarized in Table I. We also include the results of the calculations with PBE XC functional for comparison. We can see clearly from Table I that the lattice constant '$A$' obtained by these two XC functionals, namely PBE and vdW-DF, match very well with each other. The difference between these two values is about 0.2 $\%$. The lattice constant '$A$' is  determined by the strong intra-layer covalent bonds between the atom in basal plane. Both these functionals accurately describe the covalent bonds present within the layers.  However, we observe a large difference of about 7 $\%$ in value of lattice constant along $c$ axis obtained by DFT calculations with PBE and vdW-DF XC functionals. The PBE XC functional  underestimates the strength of vdW potential between the layers and hence it overestimates the value of lattice constant '$C$'. The failure of LDA and GGA is due to fact that for large inter-layer distance '$D$', the vdW attractive potential scales as -$C_4/D^4$, where $C_4$ is vdW interaction coefficient, whereas both LDA and GGA XC functionals predict the wrong trend of exponential decay with the inter-layer distance\cite{vdw-df4}. It is interesting to note that the inter-layer distance in bulk hybrid system is 3.426 $\AA$ which is comparable with 3.354 $\AA$ of bulk graphite.  Thus, our calculations on the hybrid system show that vdW interaction play a crucial role in the coupling between the layers of silicene and BN. This vdW interaction is similar to that between the layers of graphite as well as multi-layers of graphene.  

 \begin{figure}
 \begin{center}
 \includegraphics[width=6.5cm]{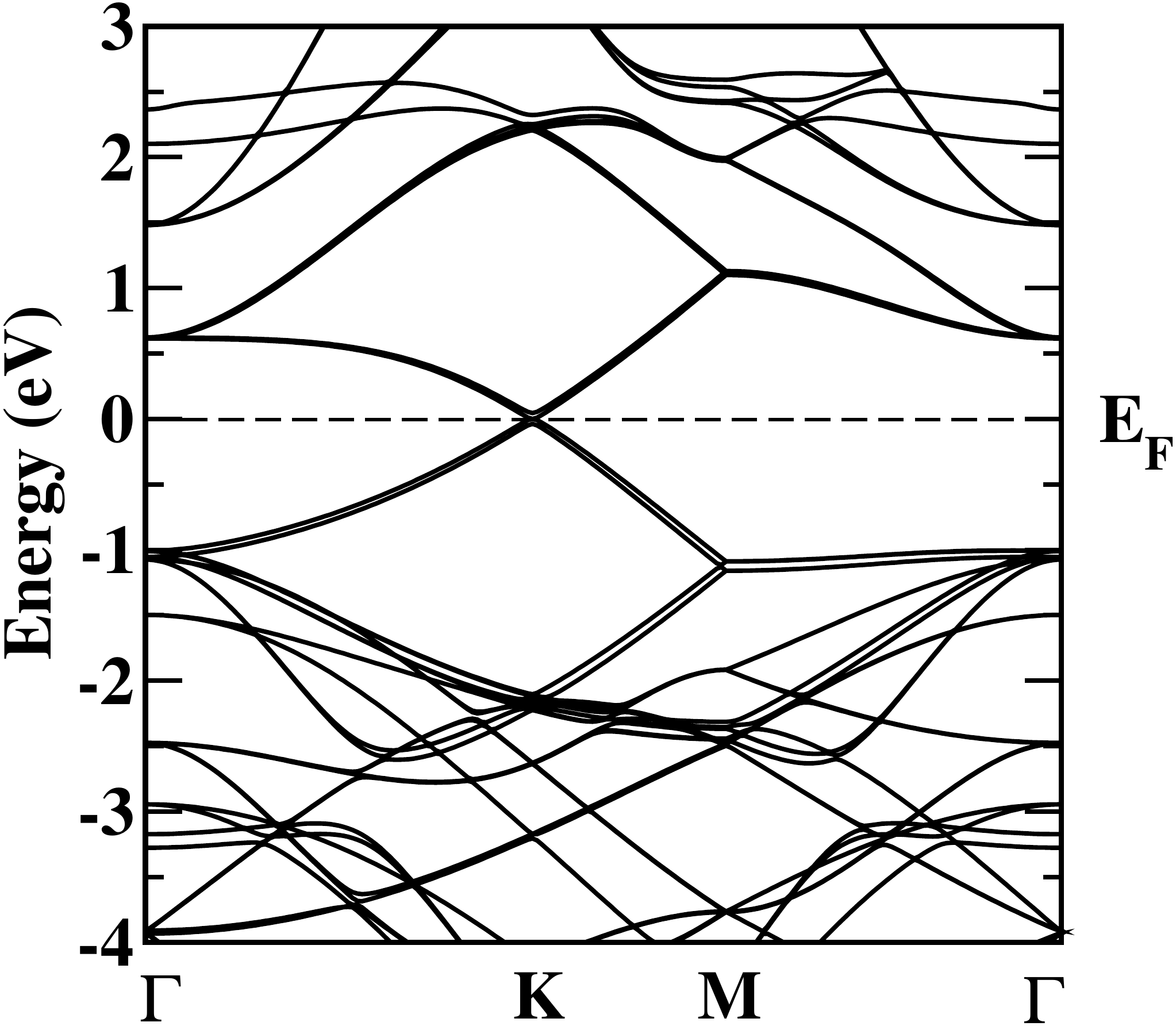}
  \caption{ Band structures of  hybrid graphite-like bi-layered structure (Si$_{16}$B$_{9}$N$_{9}$) obtained by vdW-DF. The values of energy of bands are with respect to the Fermi level.}
 \end{center}
  \end{figure}

In order to check the effect of weak vdW interaction on the electronic properties of the hybrid system, we also calculate the electronic band structures for the hybrid bilayer system by employing DFT with vdW-DF XC functional. The results of these calculations are given in Fig. 8. We compare these results with the corresponding one obtained by using PBE XC functional (See Fig. 2 (e)). We observe that the two functionals give similar results for the electronic band structure along $\Gamma$-K-M-$\Gamma$ path in Brillouin zone of the hybrid system and thus the effect of vdW interaction on the electronic properties is negligible.

\section{Acknowledgments}
 Authors thank Dr. P. D. Gupta, Dr. S. K. Deb and Dr. P. K. Gupta for encouragement and support. CK thanks Mr. Himanshu Srivastava for useful discussions. The support and help of Mr. P. Thander and the scientific computing group, Computer Centre, RRCAT is acknowledged.

\pagebreak

\end{document}